\documentclass[reprint,prb]{revtex4-1}
\usepackage{bm,epsfig,graphicx,color,amssymb,amsmath}
\begin{document}
\title{Exciton-plasmaritons in graphene-semiconductor structures}

\author{Kirill A. Velizhanin}

\email{kirill@lanl.gov}

\affiliation{Theoretical Division, Los Alamos National Laboratory, Los Alamos, New Mexico 87545, USA}

\author{Tigran V. Shahbazyan}

\email{shahbazyan@jsums.edu}

\affiliation{Department of Physics, Jackson State University, Jackson, Mississippi 39217, USA}


\begin{abstract}
We study strong coupling between plasmons in monolayer charge-doped graphene and excitons in a narrow gap semiconductor quantum well separated from graphene by a potential barrier. We show that the Coulomb interaction between excitons and plasmons result in mixed states described by a Hamiltonian similar to that for exciton-polaritons and derive the exciton-plasmon coupling constant that depends on system parameters. We calculate numerically the Rabi splitting of exciton-plasmariton dispersion branches for several semiconductor materials and find that it can reach values of up to  50 - 100 meV. 
\end{abstract}

\maketitle

\section{Introduction}

Graphene plasmonics  has recently emerged as a promising platform for studying strong light-matter interactions.\cite{Grigorenko2012-749}  Graphene is  a novel two-dimensional (2D) material \cite{Novoselov2004-666,Zhang2005-201} with unique electronic and optical properties \cite{Geim2007-183,CastroNeto2009-109} and a wide range of applications such as biosensors,\cite{Shao2010-1027} ultrafast lasers, \cite{Sun2010-803} optical modulators, \cite{Liu2011-64} and photodetectors. \cite{Fang2012-3808} Clean graphene samples are characterized by long electron scattering times and much lower, compared to metals, Ohmic losses due to relatively weak electron-phonon interaction. \cite{Chen2007-206} Charge-doped graphene with Fermi energy, $E_{F}$, in the range 0.2-0.6 eV exhibits a stable in-plane plasmon  in the infrared frequency range \cite{Fei2011-4701,Yan2012-330,Chen2012-77,Fei2012-82,Zhu2013-131101,Despoja2013-075447} with gate-tunable wavelength, $\lambda_{p}$, well below radiation wavelength at the same frequency. \cite{Jablan2009-245435} 

A large local density of states (LDOS) of graphene plasmons as compared to that of surface plasmon polaritons (SPP) on metal surfaces ensures very efficient plasmon excitation by a local probe such as  an excited dye molecule or an exciton in semiconductor quantum dot (QD) situated at a close distance  to the graphene sheet. \cite{Koppens2011-3370,Velizhanin2011-085401,Nikitin2011-195446,Gomez-Santos2011-165438} Recent measurements  \cite{Gauderau2013-2030} of nonradiative energy transfer from a dye molecule to graphene pointed to a strong Coulomb coupling between  electronic excitations in these two systems with transfer rates exceeding the radiative decay rate by a factor of $\sim$10$^{2}$. Plasmons can also provide efficient coupling between several emitters situated near a graphene sheet, e.g., by enhancing superradiance from two excited QDs \cite{Huidobro2012-155438} or facilitating long-distance energy transfer between a donor and an acceptor. \cite{Velizhanin2012-245432} These studies explored weak coupling regime between graphene plasmons and excitons, i.e., the energy spectra of these excitations are not significantly altered by the optical interactions between them.

Here we demonstrate that a \textit{strong} coupling between excitons and graphene plasmons can be realized in a hybrid graphene-semiconductor system giving rise to a propagating exciton-plasmon state with a mixed  2D dispersion (hereafter referred to as exciton-plasmariton\cite{shah-prb71}). Strong exciton-plasmon coupling effects have been recently studied between surface plasmon-polaritons (SPP) or localized surface plasmons in metal structures and excitons in dye J-aggregates, \cite{Bellessa2004-036404,Sugawara2006-266808,Wurtz2007-1297,
Fofang2008-3481,Bellessa2009-033303,Fofang2011-1556,Guebrou2012-066401,
Schlather2013-3281} individual dye molecules, \cite{Hakala2009-053602,Berrier2011-6226,Salomon2012-073002} semiconductor quantum wells (QW), \cite{Vasa2008-116801,Lawrie2012-6152} or QDs and nanoscrystals. \cite{Gomez2010-274,Gomez2013-4340,Manjavacas2011-2318} In these systems, very large Rabi splittings (relative to vacuum Rabi splitting \cite{Khitrova2006-81}) were reported  in the range of 100-200 meV for  dispersion of SPP coupled to molecular excitons  \cite{Bellessa2004-036404,Hakala2009-053602,Berrier2011-6226,
Salomon2012-073002} and  in the range of 200-450 meV  for absorption/transmission spectra of excitons coupled to localized plasmons (plexcitons), \cite{Sugawara2006-266808,Wurtz2007-1297,Fofang2008-3481,
Bellessa2009-033303,Schlather2013-3281,Fofang2011-1556,Lawrie2012-6152} while relatively small splittings ($<10$ meV) were observed for SPP coupled to QW excitons.\cite{Vasa2008-116801,Lawrie2012-6152} The optical excitation energies in the above experiments were in the range 1-3 eV to match those of plasmon resonances in metal nanostructures. In contrast, stable graphene plasmons have significantly lower energies (below $0.5$ eV in highly doped samples) and, to the best of our knowledge, strong coupling regime in graphene has not yet been explored. Here we show that a strong exciton-graphene plasmon coupling can be achieved in a hybrid structure comprised of doped graphene monolayer separated by a thin spacer from a narrow gap semiconductor QW.

\begin{figure}
\includegraphics[width=3in]{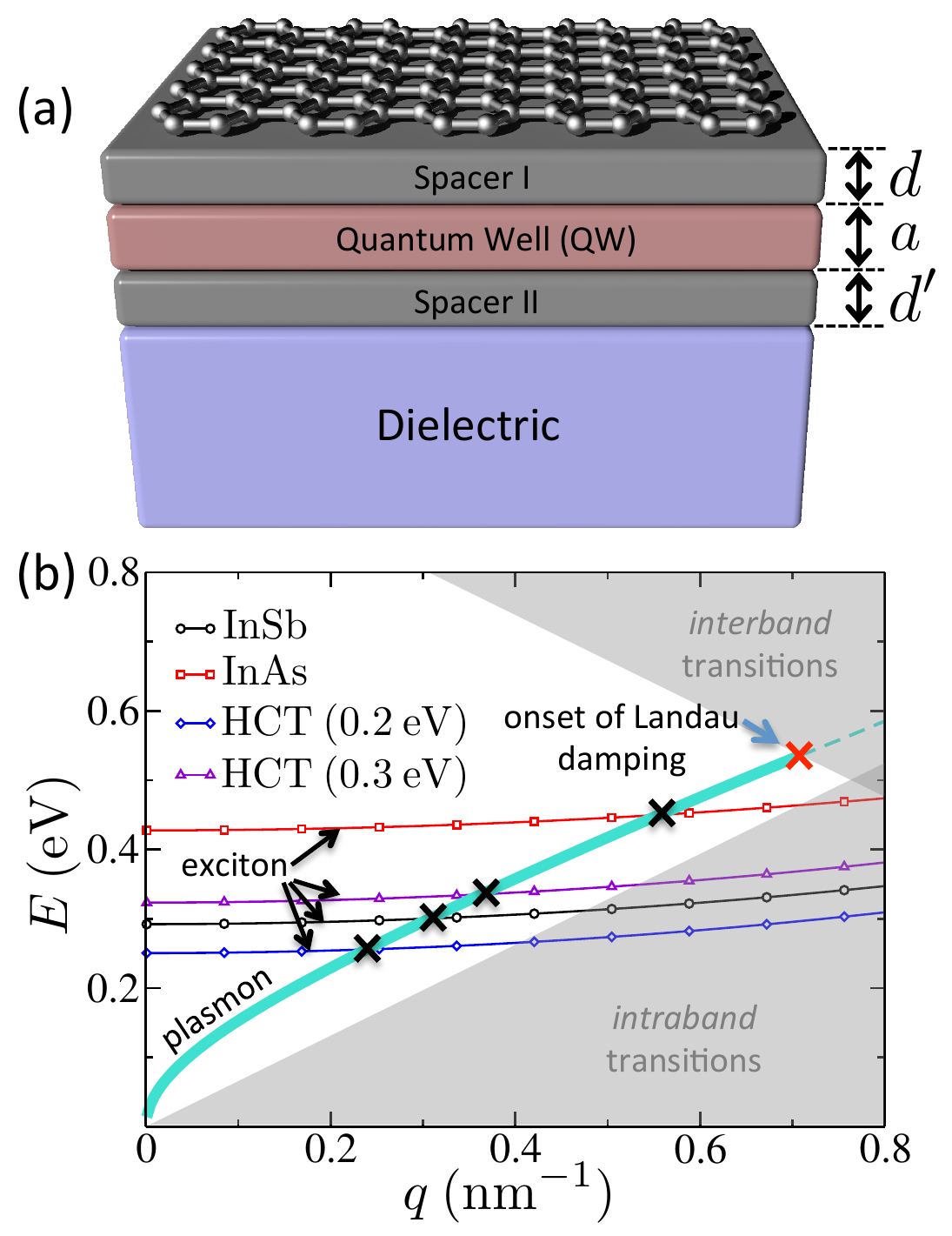} 
\caption{\label{fig:Fig1} 
(a) Schematics of the device. (b) Plasmon and exciton dispersion relations are shown by thick turquoise and thin lines, respectively. The bulk bandgap of HCT is given in  parentheses in the legend. The QW width and the Fermi energy in graphene are $a=20$ nm and $E_F=0.5$ eV, respectively. Plasmon-exciton resonances and the onset of Landau damping are marked by black and red crosses, respectively. Graphene electron-hole transition continua are shown in grey.
}
\end{figure}

The proposed structure is schematically shown in Fig.~\ref{fig:Fig1}(a). The tunneling barrier between QW and graphene, Spacer I, prevents a photoinduced charge transfer from QW to graphene. In this case, quasiparticles in graphene (plasmons) and QW (excitons) interact with each other only via direct Coulomb coupling. The QW rests on a dielectric substrate separated by Spacer II. As examples, we chose four narrow gap  semiconductor QWs: InAs, InSb, and HgCdTe (HCT) with two different Cd concentrations corresponding to HCT bulk bandgaps of $E_{g}=0.2$ eV and 0.3 eV. \cite{Becker2011} In Fig.~\ref{fig:Fig1}(b), the exciton dispersion relations for all  QWs are plotted together with the plasmon dispersion. In principle, both exciton and plasmon energies  as well as the Coulomb coupling between them are subject to complex dielectric screening determined by spacer layer widths $d$ and $d'$, QW width $a$, and dielectric constants of all the materials involved. For example, by varying the parameters of the layered structure, the Coulomb coupling strength can be tuned in a wide range. In this paper, we assume a simplified model of dielectric screening by introducing a single effective dielectric constant, $\kappa$, and using it as a free parameter. For example, Fig.~\ref{fig:Fig1}(b) shows the graphene plasmon and QW exciton dispersions for effective dielectric constant of $\kappa=7$  and the graphene charge-doping level of $E_F=0.5$ eV. At this doping level, graphene plasmon is Landau damped at $q\gtrsim 0.7$ nm$^{-1}$, the onset of damping marked by red cross in Fig.~\ref{fig:Fig1}(b). It can be seen that, for the chosen semiconductor QWs, the plasmon and excitons dispersion curves intersect at $q$ well below the damping onset. At smaller $\kappa$, exciton energies decrease (due to stronger electron-hole interaction) and plasmon energies increase, resulting in the exciton-plasmon resonance at even smaller $q$. For small separations between graphene and QW, we thus expect a mixed  exciton-plasmon state to form due to strong Coulomb coupling between its constituent excitations. 

We show that the dynamics of exciton-plasmaritons can be accurately described by a simple Hamiltonian similar to exciton-polariton Hamiltonian with exciton-plasmon coupling constant strongly affected by the  system parameters.  We explore the strong coupling regime by performing numerical calculations involving full graphene response functions  and compare the results to our model. The calculated Rabi splitting between the exciton-plasmariton upper and lower dispersion branches strongly depends on system parameters and can reach values $\sim$50-100 meV for small graphene-QW separation $d$.

The paper is organized as follows. In Sec. \ref{sec:hamiltonian} we present our model for coupled QW exciton and graphene plasmon and derive the exciton-plasmariton Hamiltonian and its energy spectrum within plasmon-pole approximation. In Sec. \ref{sec:numerics} we perform numerical calculations of exciton-plasmariton energy spectrum using graphene full response functions and compare the results to those of our model. Technical details of calculations are described in Appendices.

\section{Exciton-plasmariton Hamiltonian}
\label{sec:hamiltonian}

We consider a QW of width $a$ separated from a monolayer graphene by a sufficiently high potential barrier of thickness $d$. The system Hamiltonian is $H=H_{\rm QW}+H_{\rm G}+H_{\rm QW\mbox{-}G}$, where $H_{\rm QW}$ and $H_{\rm G}$ are many-body Hamiltonians for electronic excitations in QW and graphene, respectively, and ${H}_{\rm QW\mbox{-}G}$ describes the Coulomb interaction between them. In the absence of free electron population in QW, the main contribution to $H_{\rm QW\mbox{-}G}$ comes from the interaction of interband polarization in QW with the electric field due to charge density fluctuations in graphene (see Appendix~\ref{app:Ham})
\begin{equation}
H_{QW-G}=-\int dV\: \textbf{d}_{cv}\cdot \textbf{E} + H.c.,
\end{equation}
where $\textbf{d}_{cv}=\psi_{c}^{*}(\textbf{r})e\textbf{r}\psi_{v}(\textbf{r})$ is the dipole transition operator between conduction ($c$) and valence ($v$) bands and $\psi_{i}$ are Bloch functions ($i=c,v$).\cite{Haug2009} The electric field $\textbf{E}(\textbf{r})=-e^{-1}\nabla \Phi(\textbf{r})$ acting on exciton in QW is generated by density fluctuations in graphene
 \begin{equation}
 \label{potential}
\Phi(\textbf{r})=\int_{G} d\textbf{r}'_{\parallel}\:V(\textbf{r}-\textbf{r}')\left [\rho^{\rm G}(\textbf{r}'_{\parallel})-\rho_{0}^{\rm G}\right ],
\end{equation}
where $\textbf{r}=(\textbf{r}_{\parallel},z)$ is 3D coordinate, $\textbf{r}_{\parallel}$ and $z$ being in-plane and out-of-plane coordinates, respectively, and integration is carried over graphene plane $z=0$. The in-plane Fourier transform of the Coulomb potential, $V({\bf r})=1/\kappa r$, is given by $V(\textbf{q},z)=\frac{2\pi e^{2}}{{\kappa} q}e^{-qz}$, where $\textbf{q}$ is the in-plane momentum and ${\kappa}$ is the effective dielectric constant. Electron density operator and the average electron density in graphene are denoted by $\rho^{\rm G}(\textbf{r}_{\parallel})$ and $\rho_{0}^{\rm G}$, respectively, .

Our goal is to derive effective exciton-plasmon Hamiltonian. To this end, let us first recast $H_{\rm QW\mbox{-}G}$ in second-quantized electron-hole form. In a standard manner,\cite{Haug2009} the Bloch function splits into envelope and periodic parts, $\psi_{i\bf k}=\varphi_{i\bf k}(r)u_{i\bf k}(r)$, where the latter defines the interband dipole matrix element averaged over unit cell volume ${\cal V}$, ${\bf r}_{cv}={\cal V}^{-1}\int_{cell} d{\bf r}\: u^*_{c\bf k}({\bf r}){\bf r}u_{v\bf k}({\bf r})$, while the envelope functions $\varphi_{i\bf k}({\bf r})=S^{-1/2} e^{i{\bf k}\cdot{\bf r}_\parallel}f_i(z)$ describe "free" electrons in QW, where $f_i(z)$ is the size-quantization wave function in QW and $S$ is the normalization area (hereafter set to unity). Performing 2D Fourier transform, we obtain
\begin{equation}
H_{\rm QW\mbox{-}G}=\sum_{{\bf k}{\bf q}} h^{cv}_{{\bf q}}a^\dagger_{{\bf k}-{\bf q}}b^\dagger_{-{\bf k}}\rho^{\rm G}_{\bf q}+H.c.,
\label{eq:Hint_eh}
\end{equation}
where $a_{\textbf{k}}\equiv a_{c\textbf{k}}$, $b_{-\textbf{k}}^{\dagger} \equiv  a_{v\textbf{k}}$ are electron (annihilation) and hole (creation) operators defined in a usual way through their conduction and valence band counterparts,\cite{Haug2009} and $\rho^{\rm G}_{\bf q}=\int d\textbf{r}_{\parallel}\:e^{-i\textbf{q}\cdot \textbf{r}_{\parallel}}\varphi^{\rm G\dagger}(\textbf{r}_{\parallel})\varphi^{\rm G}(\textbf{r}_{\parallel})$ is the Fourier transform of graphene density operator, $\varphi^{\rm G}(\textbf{r}_{\parallel})$ being electron operator in graphene (as usual, the point $\textbf{q}=0$ is excluded). The matrix element $h^{cv}_{{\bf q}}$ is evaluated as 
\begin{equation}
h^{cv}_{{\bf q}}=\frac{2\pi e^{2}}{{q\kappa}}\left[i\textbf{q}\cdot\textbf{r}_{cv}-q\hat{\textbf{z}}\cdot \textbf{r}_{cv}\right]u_{cv}(q),
\end{equation}
where (for $z>0$)
\begin{equation}
u_{cv}(q)=e^{-qd}\int_0^a dz\: f_c^*(z)f_v(z)e^{-qz}
\end{equation}
is overlap between gpaphene Coulomb potential and QW excitations across the structure. We now introduce the exciton creation operator in a standard manner \cite{Haug2009} as 
\begin{equation}
c_{\nu \textbf{q}}^{\dagger}
=\sum_{\textbf{p}}{\phi}_{\nu}(\textbf{p}) a_{\textbf{p}+\alpha_{e}\textbf{q}}^{\dagger} b_{-\textbf{p}+\alpha_{h}\textbf{q}}^{\dagger},
\end{equation}
where  $\alpha_{e}=m_{e}/M$, $\alpha_{h}=m_{h}/M$ ($M=m_{e}+m_{h}$) are the relative electron and hole effective masses, and ${\phi}_{\nu}(\textbf{p})$ is the 2D Fourier transform of eigenfunction $\phi_{\nu}(\textbf{r})$ of the Wannier equation (${\bf q}$ is the exciton center of mass momentum and $\nu$ is its quantum state). Using  the orthogonality relation $\sum_{\nu}{\phi}_{\nu}(\textbf{p}){\phi}_{\nu}^{*}(\textbf{p}')=\delta_{\textbf{p}\textbf{p}'}$, Eq.~(\ref{eq:Hint_eh}) takes the form
\begin{equation}
\label{ham-int}
H_{\rm QW\mbox{-}G}=
\sum_{\nu, {\bf q}}t_{\nu \textbf{q}}c_{\nu\textbf{q}}^{\dagger}\rho^{\rm G}_{\bf q}+H.c. ,
\end{equation}
where
\begin{equation}\label{eq:t_coupl}
t_{\nu \textbf{q}}=\frac{2\pi e^{2}}{\kappa}\left (ir^{\bf q}_{cv}-r^z_{cv}\right )u_{cv}(q)\phi^*_{\nu}({\bf r}=0)
\end{equation}
characterizes the strength of exciton coupling to charge density fluctuations in graphene, and $r^z_{cv}$ and $r_{cv}^{\bf q}$ are, respectively, projections of ${\bf r}_{cv}$ onto $z$-axis and vector ${\bf q}$ (note that  $t_{\nu \textbf{q}}=t_{\nu,- \textbf{q}}^{*}$).

Other  terms of the Hamiltonian that contain QW electron and hole operators can be expressed in excitonic variables using standard projection technique.  The free exciton Hamiltonian has the usual form \cite{Haug2009} $H_{\rm QW}=\sum_{\nu q}E_{\nu q}c_{\nu\textbf{q}}^{\dagger}c_{\nu  \textbf{q}}$,
where $E_{\nu q}$ is exciton dispersion (see below) while higher-order terms including exciton-exciton scattering and exciton scattering by graphene density fluctuations are negligible for low exciton  densities we consider (see Appendix~\ref{app:Ham}). The Hamiltonian of QW exciton interacting with graphene density fluctuations then takes the form
\begin{equation}
\label{ham2}
H=\sum_{\nu, \textbf{q}}E_{\nu q}c_{\nu\textbf{q}}^{\dagger}c_{\nu  \textbf{q}} 
+\sum_{\nu, \textbf{q}} \left (t_{\nu q}c_{\nu,\textbf{q}}^{\dagger}\rho^{\rm G}_{\textbf{q}}+H.c.\right ) +H_{G}.
\end{equation}
The energy spectrum of exciton-plasmatiton can be obtained from exciton Green function "dressed" due to the interactions with graphene density fluctuations. In the rotating wave approximation, the non-interacting exciton Green function can be written as $D^0_\nu(q,\omega)=(\hbar\omega-E_{\nu q}+i\gamma_0)^{-1}$, where $\gamma_{0}$ is free exciton damping. The interaction with graphene excitations results in the appearance of the exciton self-energy in the Green function,
\begin{equation}
\label{exciton-green}
D_{\nu}({\bf q},\omega)=\frac{1}{\hbar\omega-E_{\nu q}+i\gamma_0-\Sigma_{\nu}({\bf q},\omega)},
\end{equation}
where, in the lowest order in QW-graphene coupling, the spectrum is determined by the density-density correlation function in graphene, $\Pi(q,\omega)$: $\Sigma_{\nu}({\bf q},\omega)=\left |t_{\nu {\bf q}}\right |^{2}\Pi(q,\omega)$. It is important to note here that even though the spectra of electronic excitations in graphene and QW are assumed isotropic by themselves, this is not necessarily the case for exciton-plasmaritons due to the possible anisotropy in the coupling constant, $t_{\nu{\bf q}}$. This anisotropy originates from the directionality of ${\bf r}_{cv}$ and is, therefore, maximized when ${\bf r}_{cv}$ is parallel to the graphene plane [see Eq.~(\ref{eq:t_coupl})].

Within the random phase approximation (RPA), the density-density correlation function in graphene is given by $\Pi(q,\omega)=\Pi_0(q,\omega)/\left [1-{v}_{q}\Pi_0(q,\omega)\right ]$, where $\Pi_0(q,\omega)$ is the polarization bubble in graphene and ${v}_{q}=2\pi e^{2}/{\kappa}q$. 

In the following we restrict ourselves to  the lowest $1s$ excitonic state in a narrow QW ($a\ll a_{B}$) characterized by wave function $\phi({\bf r}=0)=\sqrt{\frac{2}{\pi}} a_{B}^{-1}$ where $a_B=\frac{\hbar^{2}\kappa}{2\mu e^{2}}$ is a 2D Bohr radius. The exciton energy is $E_{0q}=E_{0}+q^{2}/2M$, where
\begin{equation}
E_{0}=E_{g}+\frac{\pi^2}{2\mu a^2}-E_{B}
\end{equation}
is the excitation energy of the lowest QW exciton. Here, $E_g$ is the bulk bandgap energy, $\mu^{-1}=m^{-1}_e+m^{-1}_h$ is exciton reduced mass, $E_B=e^{2}/\kappa a_B$ is exciton binding energy, and the second term in the r.h.s. is correction to the bulk bandgap energy due to the quantum confinement in QW. For a symmetric QW, the transverse part of envelope functions of the lowest QW subband is $f_{c}(z)=f_v(z)=(2/a)^{1/2}\sin(\pi z/a)$. Then the Coulomb overlap $u_{cv}(q)$ can be explicitly evaluated as
\begin{equation}
u_{cv}(q)=e^{-qd}\,\frac{4\pi^2(1-e^{-qa})}{4\pi^2aq+a^3q^3}.\label{eq:ucv}
\end{equation}
In the long-wave limit, the graphene polarization bubble has the form $\Pi_0(q,\omega)=\frac{E_F}{\pi}\frac{q^2}{\omega^2}$, so that $\Pi(q,\omega)$ is dominated by the plasmon pole 
\begin{equation}
\Pi(q,\omega)=\frac{\Lambda_{q}}{\hbar\omega-\hbar\omega_{q}+i\gamma}, 
~~
\Lambda_{q}=\frac{\hbar^3\omega^3_{q}}{8\pi E_{F}}\left (\frac{\kappa}{e^{2}}\right )^{2},
\label{eq:Pi_PlPole}
\end{equation}
where $\hbar\omega_{q}=\sqrt{2E_{F}qe^{2}/{\kappa}}$ and $\gamma$ are, respectively, plasmon energy dispersion and damping rate (in energy units). Combining Eqs.~(\ref{eq:t_coupl}) and (\ref{eq:Pi_PlPole}), we obtain the exciton self-energy as 
\begin{equation}
\Sigma({\bf q},\omega)=\frac{|g_{\bf q}|^{2}}{\hbar\omega-\hbar\omega_{q}+i\gamma},
\end{equation}
where the  exciton-plasmon coupling, $g_{\textbf{q}}=t_{\textbf{q}}\Lambda_{q}^{1/2}$, is given by
\begin{equation}
\label{coupling}
g_{\textbf{q}}=\frac{\left (\hbar \omega_{q}\right )^{3/2}}{a_{B}E_{F}^{1/2}}\,u_{cv}(q)\left (ir^{\bf q}_{cv}-r^z_{cv}\right ).
\end{equation}
With the above self-energy, the poles of Eq.~(\ref{exciton-green}) determine two exciton-plasmariton energy branches,
\begin{equation}
\label{spectrum}
\hbar\omega_{\bf q}^{\pm}=\frac{1}{2}\left [ E'_{q}+\hbar\omega'_{q}\pm\sqrt{\left (E'_{q}-\hbar\omega'_{q}\right)^{2}+4|g_{\bf q}|^{2}}\right ],
\end{equation}
where $E'_q=E_q+i\gamma_0$ and $\hbar\omega'_q=\hbar\omega_q+i\gamma$. To gain more insight, we note that  Eq.~(\ref{spectrum}) is the energy spectrum of a simple two-level Hamiltonain where both the energies of the two levels, $E'_q$ and $\hbar\omega'_q$, and the coupling between them, $g_{\bf q}$, are dependent on ${\bf q}$. This observation can be made precise by casting the graphene Hamiltonian in terms of plasmon normal modes, $H_{\rm G}=\sum_{\textbf{q}}\hbar\omega'_{q}a_{\textbf{q}}^{\dagger}a_{\textbf{q}}$, where $a_{\textbf{q}}^{\dagger}$ is a plasmon creation operator. After expanding the graphene charge density operator in Eq.~(\ref{ham-int}) over these normal modes, \cite{Kato1999-235} the Hamiltonian (\ref{ham2}) of QW excitons interacting with graphene plasmons takes the form 
\begin{align}
\label{ham-final}
H=\sum_{\textbf{q}}\biggl [E'_{q}c_{\textbf{q}}^{\dagger}c_{\textbf{q}} +\hbar\omega'_{q}a_{\textbf{q}}^{\dagger}a_{\textbf{q}} 
~~~~~~~~~~~~~~
\nonumber\\
+ \left [g_{\bf q}c_{ \textbf{q}}^{\dagger} \left (a_{\textbf{q}}+a_{-\textbf{q}}^{\dagger}\right )  + {\rm H.c}\right ]\biggr ],
\end{align}
where  exciton-plasmon coupling $g_{\textbf{q}}$ is defined by Eq.~(\ref{coupling}). The imaginary part of $E'_q$ and $\hbar\omega'_q$ has to be interpreted in a usual manner as a decay constant of a metastable state. The corresponding lifetimes of an exciton and plasmon are given by $\tau_{0}=2\hbar/{\rm Im} E'_q$  and $\tau=2/{\rm Im} \omega'_q$, respectively. The Hamiltonian (\ref{ham-final}) is similar to that describing exciton-polaritons and can be brought to canonical form by a standard Bogolubov-Hopfield transformation \cite{Haug2009} yielding, near the resonance, two dispersion branches Eq. (\ref{spectrum}) which, in general, are anisotropic in $\textbf{q}$-plane due to the dependence of exciton-plasmon coupling $g_{\textbf{q}}$ on the exciton polarization.
Note that if ${\bf q}$ is parallel to the projection of ${\bf r}_{cv}$ onto the $xy$-plane, then the coupling is maximized and $(r^z_{cv})^2+(r^{\bf q}_{cv})^2=r^2_{cv}$, so that $|g_{\bf q}|^{2}=u_{cv}^{2}(q)r_{cv}^{2}\left (\hbar\omega_q\right )^{3}/a_B^{2}E_F$. This condition is assumed fulfilled hereafter. At resonance,  $\hbar\omega_{q_{0}}=E_{q_{0}}\approx E_{0}$, corresponding to momentum $q_{0}\approx a_{B}^{-1}\left (E_{0}^{2}/2 E_{F}E_{B}\right )$, the Rabi frequency $\Delta=\hbar\omega_{q_{0}}^{+}-\hbar\omega_{q_{0}}^{-}$ has the form
\begin{equation}
\label{rabi}
\Delta=\sqrt{4u_{cv}^{2}(q_{0})\frac{r_{cv}^{2} E_{0}^{3}}{a_B^{2} E_{F}}-(\gamma-\gamma_{0})^{2}},
\end{equation}
and strongly depends on system parameters.

All the analytical considerations above were based on the plasmon dispersion $\omega_q\propto q^{1/2}$, Eq.~(\ref{eq:Pi_PlPole}). However, this simple dispersion relation becomes inaccurate in the high-$\kappa$ dielectric environment due to non-local effects. It turns out that the dispersion relation could be analytically corrected to account for such effects (see Appendix~\ref{app:high-k} for detail). Below we compare Eq.~(\ref{spectrum}), augmented by this corrected dispersion relation, to full numerical calculations of the energy spectrum of the system.

\section{Numerical results}
\label{sec:numerics}

Numerical calculations were performed for four narrow-gap semiconductors, InSb ($E_{g}=0.235$ eV), \cite{Vurgaftman2001-5815} InAs ($E_{g}=0.4$ eV), \cite{Vurgaftman2001-5815} and HgCdTe (HCT) with two different Cd concentrations corresponding to  $E_{g}=0.2$ eV and $E_{g}=0.3$ eV. \cite{Becker2011} The electron and hole effective masses for these materials are taken from Refs.~\onlinecite{Vurgaftman2001-5815,Becker2011,Weiler1981-119,Becker2000-10353}. For each material, the dispersion of QW exciton intersects plasmon dispersion in graphene doped to $E_{F}=0.5$ eV before the Landau damping onset (see Fig.~\ref{fig:Fig1}). Unless otherwise noted, full RPA response functions for graphene were employed in numerical calculations (see Appendix~\ref{app:polariz} for detail) and standard materials parameters were used. The interband matrix element $r_{cv}$ was estimated as 
\begin{equation}
r_{cv}=|{\bf r}_{cv}|=\sqrt{\frac{E_{P}}{2E_{g}^{2}}}.
\end{equation}
where $E_{P}=2P^{2}$ is the Kane energy, $P=-i p_{cv}=E_{g}r_{cv}$ is the Kane momentum matrix element and ${\bf p}_{cv}=v^{-1}\int_{cell} d{\bf r}\: u^*_c({\bf r}){\bf p}u_v({\bf r})$ is the inter band transition momentum. For semiconductor materials considered in this work we have $E^{\rm InSb}_{P}=23.3$ eV, $E^{\rm InAs}_P=21.5$ eV and $E_P^{\rm HCT}=18.8$ eV. \cite{Vurgaftman2001-5815,Weiler1981-119,Becker2000-10353}

\begin{figure}
\includegraphics[width=3.0in]{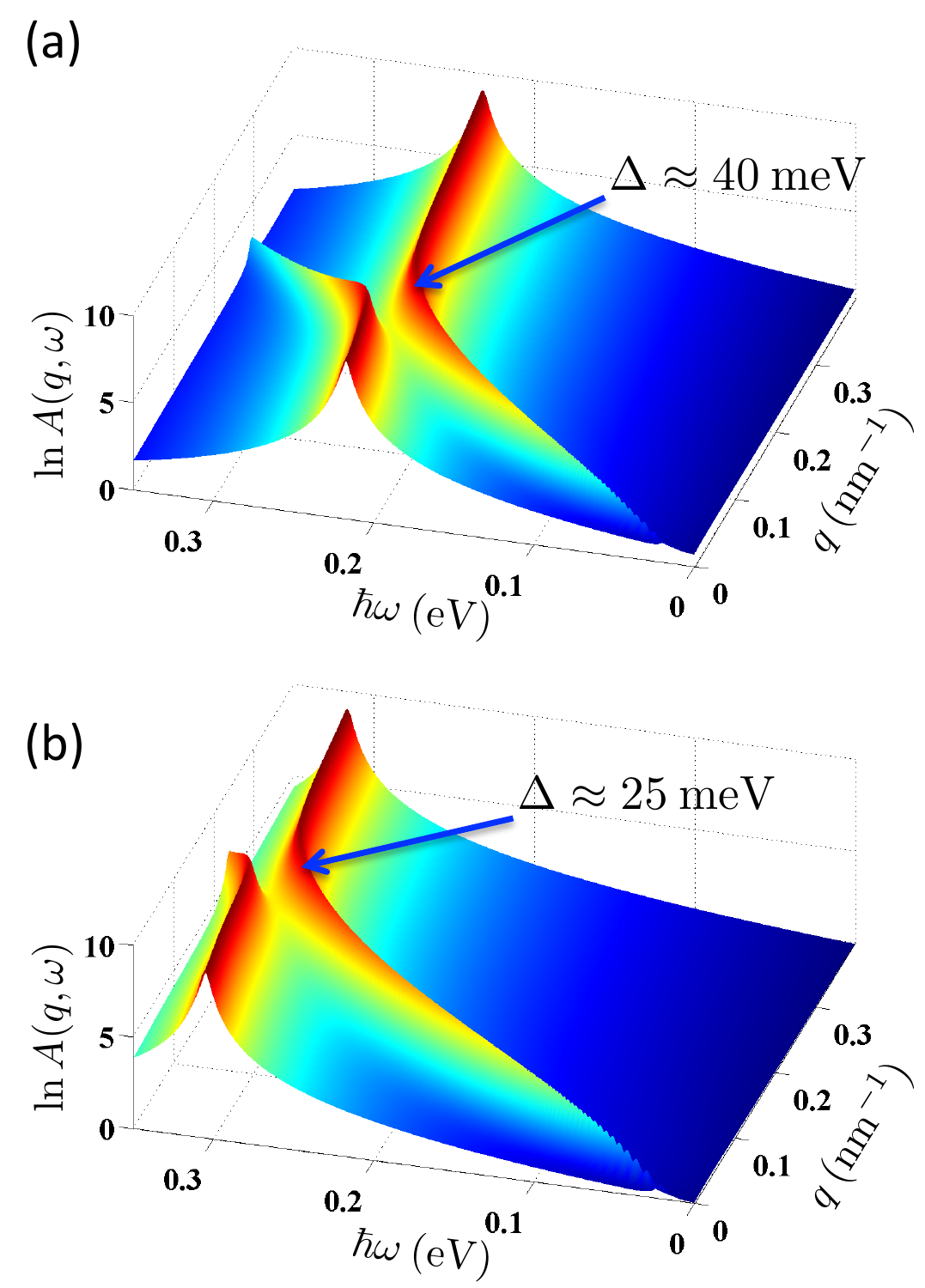} 
\caption{\label{fig:Fig2} 
Surface plot of the logarithm of spectral function for the exciton-plasmariton, $\ln A(q,\omega)$,  for (a) HCT with $E_g=0.2$ eV and (b) InAs.  The effective dielectric constant is set to $\kappa=4$ (panel a) and $\kappa=3$ (panel b). Other system parameters are set to $E_F=0.5$ eV, $d=2$ nm and $a= 20$ nm. To make the excitonic "ridge" broader and more visible, we substituted $\tau_0\rightarrow 10^{-2}\tau_0=0.1$ ps in this figure.
}
\end{figure}
Figure~\ref{fig:Fig2} shows 3D log plot of exciton spectral function $A(q,\omega)=-{\rm Im}D_0(q,\omega)$ (in arbitrary units) as defined by Eq.~(\ref{exciton-green}) for HTC and InAs QWs of width $a=20$ nm, separated from the graphene by $d=2$ nm thickness spacer, and effective dielectric constants of the structure $\kappa=4$ and $\kappa=3$, respectively. Plasmon decays time is taken $\tau=0.1$ ps. \cite{Jablan2009-245435,Koppens2011-3370} Exciton lifetime can vary in a wide range depending on specific parameters of the system, e.g., QW thickness.\cite{de-Leon2000-2874} Here we assume a very conservative value of $\tau_{0}=10$ ps. Since $\tau\ll\tau_0$, the specific value for the exciton lifetime is largely irrelevant. With the chosen parameters, both exciton-plasmariton branches are seen to be well-separated in Fig.~\ref{fig:Fig2}.

\begin{figure}
\includegraphics[width=3.0in]{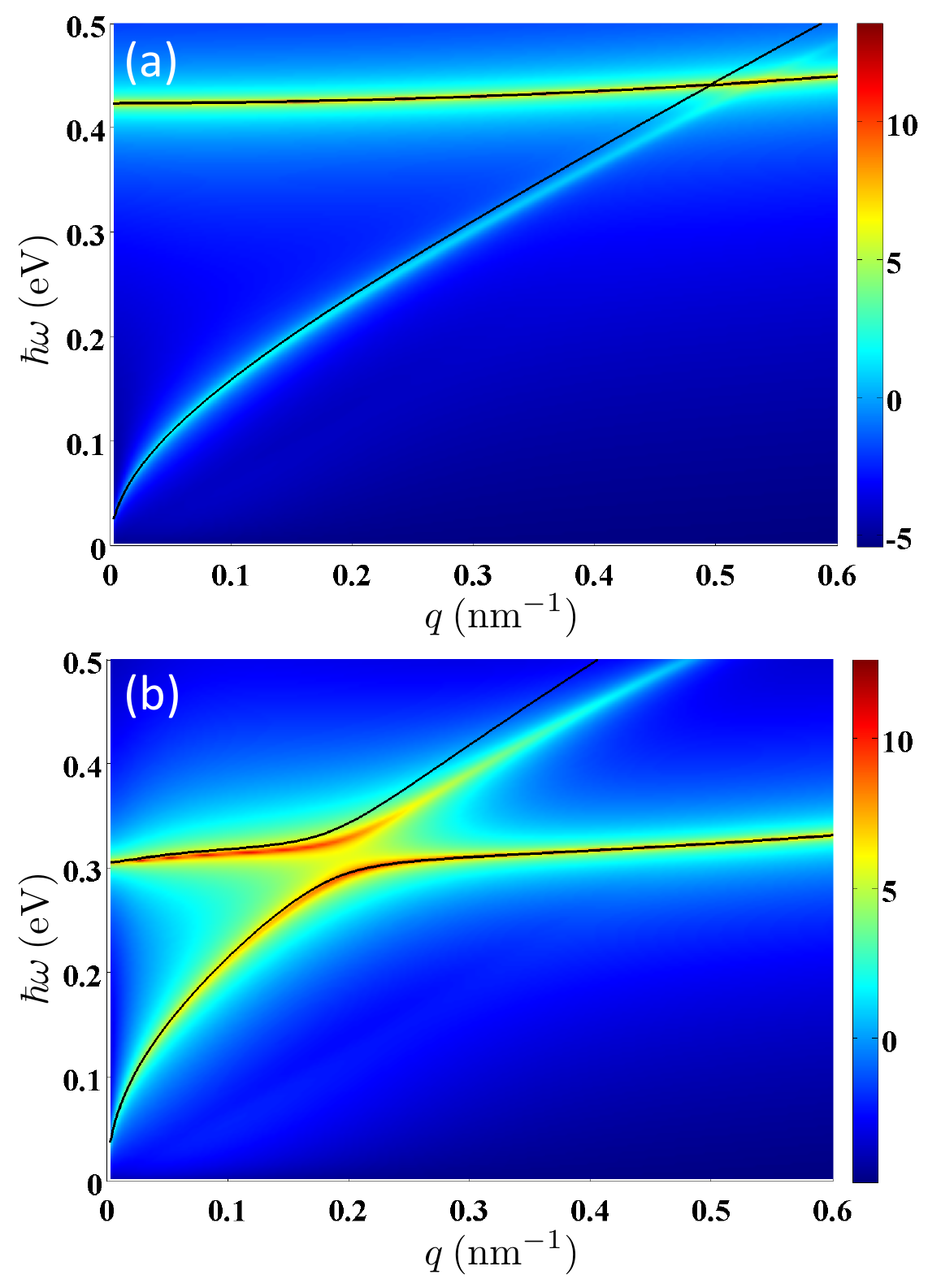} 
\caption{\label{fig:Fig3} 
Density plot of the spectral function for the exciton-plasmariton in InAs.  The effective dielectric constant is set to $\kappa=6.5$ (panel a) and to $\kappa=3$ (panel b). Other system parameters are set to $E_F=0.5$ eV, $d=2$ nm and $a= 20$ nm.
}
\end{figure}

\begin{figure}
\includegraphics[width=3.0in]{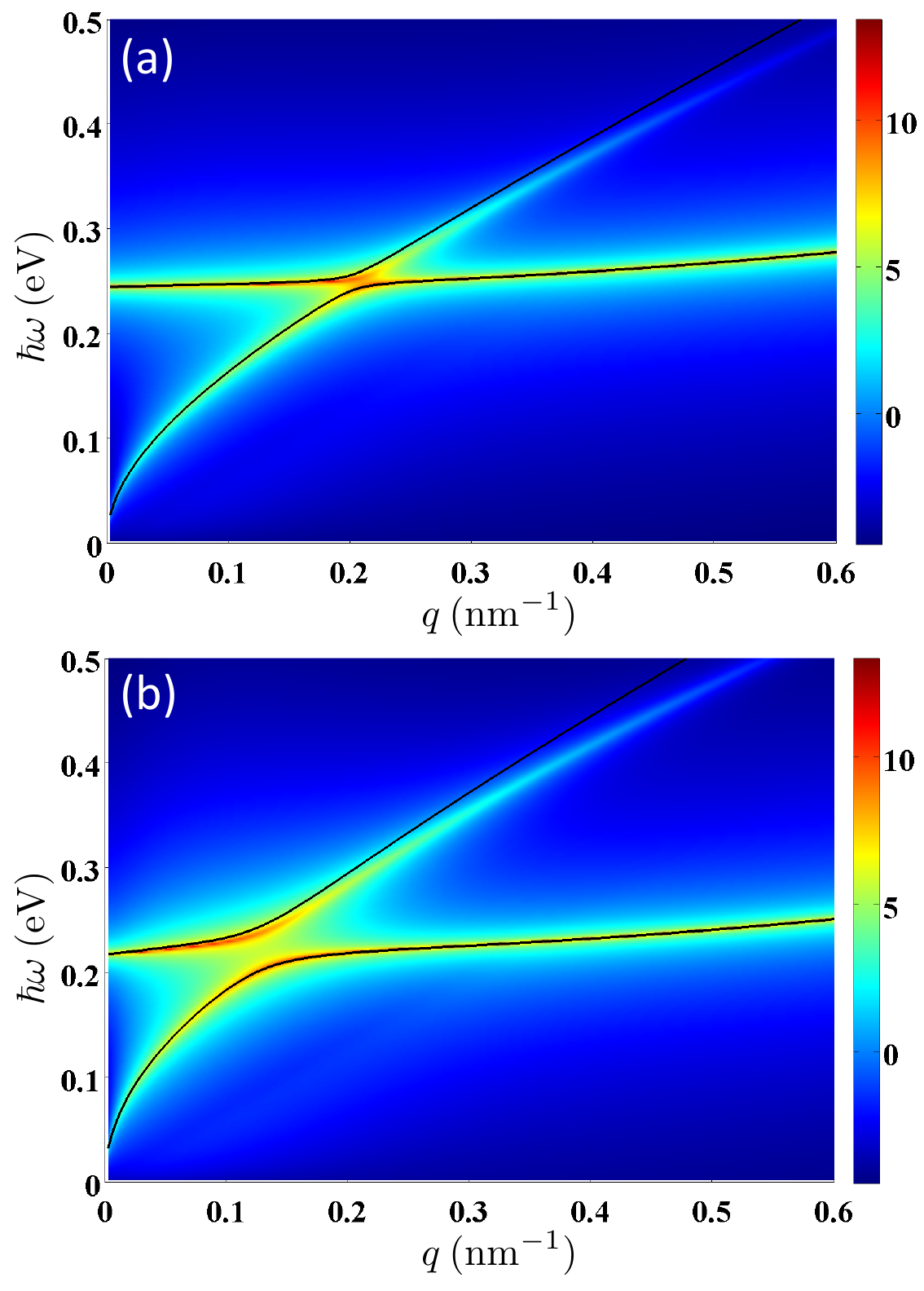} 
\caption{\label{fig:Fig4} 
Density plot of the spectral function for the exciton-plasmariton in HCT ($E_g=0.2$ eV).  The effective dielectric constant is set to $\kappa=6$ (panel a) and to $\kappa=4$ (panel b). Other system parameters are set to $E_F=0.5$ eV, $d=2$ nm and $a= 20$ nm.
}
\end{figure}

In Figs. \ref{fig:Fig3} and \ref{fig:Fig4} we show density plot of $A(q,\omega)$ vs. momentum $q$ for InAs and HTC QWs for two different effective dielectric constants. Thin black lines, superimposed on top of the density plots, show the dispersion relations of exciton-plasmariton dispersion branches calculated from Eq.~(\ref{spectrum}). We see that with increasing $\kappa$, the splitting between exciton-plasmariton branches is strongly reduced while the resonance position is shifted towards higher $q$. Both effects can be traced to exciton-plasmon coupling which is directly affected by increase of dielectric constant, $g_{q}\propto \kappa^{-1}$. At the same time, increasing $\kappa$ reduces the exciton binding energy and hence shifts the resonance position to higher $q$ which, in turn, leads to a decrease of $u_{cv}(q)$ resulting in  an even stronger reduction of $g_{q}$. 

Since the \textit{exciton} spectral function is plotted, only the exciton-like branch is expected to be seen at vanishing exciton-plasmon coupling. Indeed, the intensity of plasmon-like branches is seen to decrease for larger $\kappa$. As can be seen in all the panels, the analytical curves for the dispersion (thin black lines) agree reasonably well with the numerical results, seen as ridges in density plots. Therefore, we will rely on analytical model in our further analysis.

\begin{figure}
\includegraphics[width=3.5in]{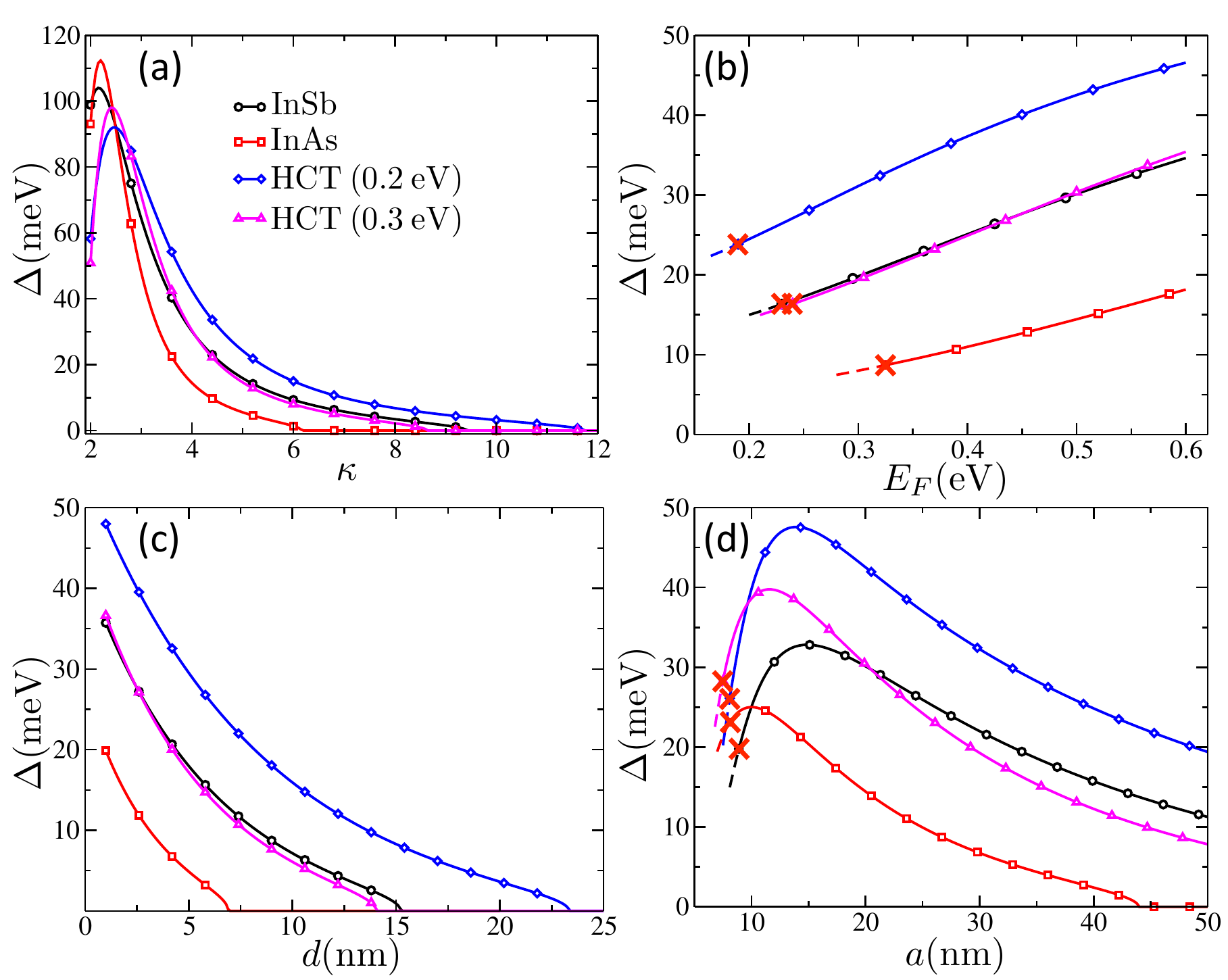} 
\caption{\label{fig:Fig5} 
Strength of exciton-plasmon coupling versus (a) effective dielectric constant, (b) Fermi energy, (c) thickness of Spacer I and (d) thickness of the quantum well. The system parameters are set to $\kappa=4$, $E_F=0.5$ eV, $d=2$ nm and $a=20$ nm except for that varied in each panel (e.g., $\kappa$ in panel a). Red crosses mark the onset of Landau damping and dashed lines correspond to short-living plasmons that live within the electron-hole sector and, therefore, are subject to Landau damping.
}
\end{figure}

In Fig.~\ref{fig:Fig5} we plot the Rabi splitting $\Delta$ given by Eq.~(\ref{rabi}) as a function of various system parameters. Panel (a) shows the dependence of Rabi splitting on the effective dielectric constant of the structure, $\kappa$. In addition to aforementioned reduction of $\Delta$ for larger $\kappa$, it has a maximum at $\kappa\sim 2.5-3$ followed by its sharp decrease for smaller $\kappa$. The latter is caused by the reduction of the {\em total} exciton energy $E_{0}$ as the exciton \textit{binding} energy $E_{B}$ increases, which leads to a sharp drop of $\Delta$ [see Eq. (\ref{rabi})]. Such a strong effect is caused by the cubic energy dependence of plasmon oscillator strength [see Eq.~(\ref{eq:Pi_PlPole})]. 

The rise of $\Delta$ as a function of Fermi energy, shown in panel (b), is due to the resonance shift to smaller momenta ($q_{0}\approx E_{0}^{2}/2E_{B}E_{F}$) which leads to increase of the Coulomb overlap $u_{cv}(q)$. It should be noted that behavior of $u_{cv}(q_{0})$ is rather complicated and depends on QW width $a$ as well as separation between graphene and QW planes $d$ [see Eq.~(\ref{eq:ucv})]. It can be seen that for InSb and HTC QWs with comparable $q_{0}$ the dependence is linear, while for InAs (larger $q_{0}$) and HTC (0.2 eV) (smaller $q_{0}$) QWs there are slight deviations from linearity. The overall magnitude is determined by the prefactor of $e^{-qd}$ originating from the graphene Coulomb potential which suppresses plasmon-exciton coupling for small $E_{F}$ corresponding to large $q_{0}$. The same prefactor determines the dependence of $\Delta$ on spacer thickness $d$ in panel (c). Indeed, the exponential decay of plasmon-exciton coupling strength with spacer thickness causes $\Delta$ to vanish for $d$ in the range from 7 nm to 24 nm for QWs used,  signaling the end of strong coupling regime. 

Finally, panel (d) shows the dependence of $\Delta$ on QW width $a$. Rabi splitting decreases as $u_{cv}(q_{0})$ in Eq.~(\ref{eq:ucv}) decreases for large $a$. For smaller $a$, however, the $a$-dependence of $\Delta$ can become non-monotonic. The reason for this is that at small $a$ the exciton energy becomes higher due to the QW size-quantization, so the plasmon and exciton dispersion intersect at higher energies (and momenta) leading to a less efficient coupling. Finally, for $a\lesssim 10$ nm the plasmon dispersion enters the electron-hole excitation continuum before it meets the exciton dispersion. In this case, the graphene plasmon is short-lived and, therefore, the excition-plasmariton is no longer a stable excitation.

\section{Conclusions}

In summary, we studied strong coupling between a graphene plasmon and a two-dimensional exciton in a narrow gap semiconductor quantum well separated from graphene by a potential barrier. We developed an analytical model based on Hamiltonian similar to that for exciton-polaritons that accurately  describes mixed exciton-plasmon states (exciton-plasmaritons) and calculated energy dispersion curves for several semiconductor materials. 

\acknowledgments Authors would like to thank G. Khodaparast for a helpful discussion. Work at LANL was performed under the NNSA of the U.S. DOE at LANL under Contract No. DE AC52-06NA25396. Work at JSU was supported by the NSF under Grant No. DMR-1206975.

\appendix
\section{Hamiltonian}\label{app:Ham}
The interaction Hamiltonian between the QW and graphene is 
\begin{equation}
\hat{H}_{QW-G}=\int_{QW} d{\bf r}\:\hat{\psi}^{\dagger}(\textbf{r})\hat{\psi} (\textbf{r})\hat{\Phi}(\textbf{r}),\label{eq:Hqwg}
\end{equation}
where $\hat{\psi} (\textbf{r})$ is the electron field annihilation operator in QW. Integral is evaluated over the volume of QW. The electrostatic potential due to charge density fluctuations in graphene is given by
\begin{equation}
\hat{\Phi}(\textbf{r})=\int_G d{\bf r}'\:V(\textbf{r}-\textbf{r}')\hat{\rho}(\textbf{r}'),
\end{equation}
where the integral is evaluated over graphene.
In the second quantization representation, electron field operators are expanded as
\begin{equation}
\hat{\psi}(\textbf{r})=\sum_{\lambda, \textbf{k}} \psi_{\lambda \textbf{k}}(\textbf{r}) \hat{a}_{\lambda \textbf{k}},
\end{equation}
where $\hat{a}_{\lambda \textbf{k}}$ is the electron annihilation operator in band $\lambda$ with (quasi)momentum $\textbf{k}$ in the first Brillouin zone and $\psi_{\lambda \textbf{k}}(\textbf{r})$ is the corresponding Bloch function. Substituting this definition of the field operator into Eq.~(\ref{eq:Hqwg}), the Hamiltonian takes the form
\begin{equation}
\label{H2}
\hat{H}_{QW-G}=\sum_{\lambda,\lambda'}\sum_{{\bf k},{\bf k}'}\hat{D} _{\lambda\lambda'}(\textbf{k},\textbf{k}')\hat{a}_{\lambda \textbf{k}}^{\dagger} \hat{a}_{\lambda' \textbf{k}'},
\end{equation}
where
\begin{equation}
\label{c}
\hat{D} _{\lambda\lambda'}(\textbf{k},\textbf{k}')=\int_{QW} d{\bf r}\: \psi^{*}_{\lambda \textbf{k}}(\textbf{r}) \psi_{\lambda' \textbf{k}'}(\textbf{r})\hat{\Phi}(\textbf{r}).
\end{equation}
In turn, Bloch functions can be expand into the basis of Wannier functions, $w_{\lambda}$, as \cite{Haug2009}
\begin{equation}
\psi_{\lambda \textbf{k}}(\textbf{r})=N^{-1/2}\sum_{n}e^{i\textbf{k}\textbf{r}_{n}}w_{\lambda}(\textbf{r}-\textbf{r}_{n}),\label{eq:Wannier}
\end{equation}
where $N$ is the total number of unit cells and the sum runs over lattice sites $\textbf{r}_{n}$. Wannier functions are normalized as $\int d{\bf r}\: w^{*}_{\lambda}(\textbf{r}-\textbf{r}_{m}) w_{\lambda'}(\textbf{r}-\textbf{r}_{n})=\delta_{\lambda\lambda'}
\delta_{mn}$. Rewriting $\hat{D}_{\lambda\lambda'}({\bf k},{\bf k}')$ via Wannier functions one obtains
\begin{align}
\label{c2}
\hat{D}_{\lambda\lambda'}(\textbf{k},\textbf{k}')&=\frac{1}{N}\sum_{m,n}  e^{i\textbf{k}'\textbf{r}_{n}-i\textbf{k}\textbf{r}_{m}}\nonumber\\ 
&\times\int d{\bf r}\: w^{*}_{\lambda}(\textbf{r}-\textbf{r}_{m}) w_{\lambda'}(\textbf{r}-\textbf{r}_{n})\hat{\Phi}(\textbf{r}).
\end{align}
Since the potential $\hat{\Phi}(\textbf{r})$ is smooth on atomic scale, we expand it within each unit cell as $\hat{\Phi}(\textbf{r})\approx \hat{\Phi}(\textbf{r}_{n}) + (\textbf{r}-\textbf{r}_{n})\cdot\nabla \hat{\Phi}(\textbf{r}_{n})$. The first term, after substituting into (\ref{c2}) yields 
\begin{equation}
\hat{D}_{\lambda\lambda'}^{(1)}({\bf k},{\bf k}')= \delta_{\lambda\lambda'}\hat{\Phi}(\textbf{k}-\textbf{k}')
\end{equation}
where $\hat{\Phi}(\textbf{k})$ is the Fourier transform of $\hat{\Phi}(\textbf{r})$. To evaluate the second expansion term, we neglect the overlap of localized Wannier wavefunctions at different lattice sites so that diagonal terms vanish. For non-diagonal terms we obtain
\begin{equation}
\hat{D} _{cv}^{(2)}({\bf k},{\bf k}')=i\textbf{d}_{cv}\cdot (\textbf{k}-\textbf{k}')\hat{\Phi}(\textbf{k}-\textbf{k}'),
\end{equation}
where $\textbf{d}_{cv}=\int_{QW}d{\bf r}\: w^{*}_{c}(\textbf{r}) {\bf r}w_{v}(\textbf{r})$ is the valence to conduction band transition dipole moment. Using the definition of Wannier functions, Eq.~(\ref{eq:Wannier}), it can be rewritten in a more standard form
\begin{equation}
{\bf d}_{cv}=\int_{u.c.} d{\bf r}\: u^*_c({\bf r}){\bf r}u_v({\bf r}),
\end{equation}
where $u_{c(v)}({\bf r})$ is the conduction (valence) bandedge Bloch function, and the integration is assumed over the unit cell.

Substituting $\hat{D}^{(1)}$ and $\hat{D}^{(2)}$ into Eq.~(\ref{H2}) we obtain $\hat{H}_{QW-G}=\hat{H}_{QW-G}^{(1)}+\hat{H}_{QW-G}^{(2)}$ with 
\begin{align}
\label{H3}
&\hat{H}_{QW-G}^{(1)}=\sum_{\lambda=c,v}\sum_{\textbf{k},\textbf{q}}\hat{\Phi}(\textbf{q})\hat{a}_{\lambda \textbf{k}+\textbf{q}}^{\dagger} \hat{a}_{\lambda \textbf{k}},
\\
&
\hat{H}_{QW-G}^{(2)}=i\sum_{ \textbf{k},\textbf{q}} (\textbf{d}_{cv}\cdot \textbf{q}) \hat{\Phi}(\textbf{q})\hat{a}_{c,\textbf{k}+\textbf{q}}^{\dagger} \hat{a}_{v,\textbf{k}}+H.c.
\end{align}
The latter expression can be rewritten in a more intuitive form
\begin{equation}
\hat{H}^{(2)}_{QW-G}=-\sum_{\bf q} \hat{\bf d}_{cv}({\bf q})\cdot\hat{\bf E}({\bf q})+H.c.,
\end{equation}
where $\hat{\bf d}_{cv}({\bf q})=\sum_{\bf k} {\bf d}_{cv}\hat{a}^\dagger_{c,{\bf k}+{\bf q}}\hat{a}_{v,{\bf k}}$ is the transition dipole moment operator of the QW unit cell. The operator of electric field due to charge density fluctuations in graphene is $\hat{\bf E}({\bf q})=-i{\bf q}\hat{\Phi}({\bf q})$. 

The Hamiltonian $\hat{H}^{(2)}_{QW-G}$ describes interaction of  {\em interband} QW excitations (excitons)  with the electric field of graphene. It represents the starting point of this work. In contrast, $\hat{H}^{(1)}_{QW-G}$ describes interactions of \emph{intraband} density fluctuations in QW with those in graphene. After recasting  in exciton basis, this term describes QW exciton scattering on graphene plasmons. At low exciton densities and at low temperatures these processes are weak and not considered here. 

\section{Density correlation function}\label{app:polariz}

The bare density correlation function (polarization bubble), is calculated within the Dirac electrons approximation as \cite{Wunsch2006-318,Hwang2007-205418,Koppens2011-3370}
\begin{gather}
\Pi_{0}(q,\omega)=  -\frac{q^2}{4\pi\hbar}\left[\frac{8E_{F}}{\hbar v_{F}^{2}q^{2}}\right.\nonumber \\
+\frac{G(-\Delta_{-})\theta\left[-{\rm Re}\left\{ \Delta_{-}\right\} -1\right]}{\sqrt{\omega^{2}-v_{F}^{2}q^{2}}}\nonumber \\
+ \left.\frac{\left[G(\Delta_{-})+i\pi\right]\theta\left[{\rm Re}\left\{ \Delta_{-}\right\} +1\right]-G(\Delta_{+})}{\sqrt{\omega^{2}-v_{F}^{2}q^{2}}}\right],\label{eq:full_pop}
\end{gather}
where $G(z)=z\sqrt{z^{2}-1}-\ln\left(z+\sqrt{z^{2}-1}\right)$ and
$\Delta_{\pm}=\left(\hbar\omega\pm2 E_{F}\right)/\hbar v_F q$.
The square roots are chosen to yield positive real parts and the imaginary
part of the logarithm is taken in $(-\pi,\pi]$ range. Fermi velocity and Fermi level (the latter determines the extent of graphene charge-doping) are denoted by $v_F$ and $E_F$, respectively. Within the Dirac electrons approximation, the density correlation function  is insensitive to the sign of the Fermi level, so in all the expressions here and in the main text $E_F$ has to be understood as $|E_F|$.

The two important limiting forms of the density correlation function are (i) the long wavelength limit ($q\rightarrow0$, $\hbar\omega\ll2 E_{F}$),  and (ii) the static limit ($\omega\rightarrow 0$, $q<2E_F/\hbar v_F$). The long wavelength limit is given by
\begin{equation}
\Pi_{0}(q\rightarrow 0,\omega)=\frac{E_{F}q^{2}}{\pi\hbar^{2}\omega^{2}}.\label{eq:pop_lowq}
\end{equation}
The static limit of the bare density correlation function is obtained as
\begin{equation}
\Pi_{0}(q,\omega\rightarrow 0)=-\frac{2E_{F}}{\pi\hbar^{2}v^2_F}.\label{eq:pop_stat}
\end{equation}
 The naive substitution $\omega\rightarrow\omega+i\gamma/2\hbar$ to account
for in-graphene scattering losses in Eq.~(\ref{eq:full_pop}) ($\gamma$ is the electron scattering
rate in energy units) is inaccurate in a general case (especially if $\gamma$
is not small), since it does not preserve the particle conservation requirement.
To correct for this, the more accurate Mermin procedure is adopted,
yielding \cite{Mermin1970-2362,Ropke1999-365}
\begin{equation}
\Pi_{\gamma}(q,\omega)=\frac{(1+i\gamma/\hbar\omega)\Pi_{0}(q,\omega+i\gamma/\hbar)}{1+(i\gamma/\hbar\omega)\Pi_{0}(q,\omega+i\gamma/\hbar)/\Pi_{0}(q,0)}. \label{eq:Mermin}
\end{equation}
The full (or ``dressed") density correlation function, which accounts for screening
in graphene, is obtained within the random phase approximation as
\begin{equation}
\Pi(q,\omega)=\frac{\Pi_{\gamma}(q,\omega)}{1-e^{2}v_q\Pi_{\gamma}(q,\omega)},\label{eq:pop_rpa}
\end{equation}
where $v_q=2\pi e^2/\kappa q$ is the two-dimensional Fourier transform of the
Coulomb potential within the graphene's plane, $v_r=e^2/\kappa r$. The effective dielectric constant of the environment $\kappa$ is in general determined by the specific geometry of the graphene-based device and bulk dielectric constants of its material constituents. For example, in the simplest case of the interface between two homogeneous materials with dielectric constants $\kappa_1$ and $\kappa_2$, the effective dielectric constant at the interface is given by $\kappa=(\kappa_{1}+\kappa_{2})/2$. \cite{Smythe1968,Ponomarenko2009-206603}

The plasmon dispersion relation, $\omega_q=\omega_p(q)$, is found by requiring the real part of the denominator of Eq.~(\ref{eq:pop_rpa}) to vanish.  The Taylor expansion of the denominator around this point (up to leading terms in both real and imaginary parts) leads to the possibility of approximating  the full density correlation function within the so called plasmon pole approximation as (for $\omega>0$)
\begin{equation}
\Pi_{pp}(q,\omega)=\frac{\Lambda_q}{\hbar\omega-\hbar\omega_q+i\Gamma_q/2},\label{eq:plpole}
\end{equation}
where $\Lambda_q=\hbar\Pi_\gamma(q,\omega_q)/A_q$ is the amplitude of plasmon pole, and $\Gamma_q=2\hbar B_q/A_q$ is the plasmon energy-dissipation rate. The coefficients of the Taylor expansion of the denominator of Eq.~(\ref{eq:pop_rpa}) are
\begin{align}
A_q&=-v_q\left.\frac{\partial}{\partial \omega}{\rm Re}\left[\Pi_{\gamma}(q,\omega)\right]
\right|_{\omega=\omega_q},\nonumber \\
B_q&=-v_q{\rm Im}\left[\Pi_{\gamma}(q,\omega_q)\right].
\end{align}
In the low-$q$ limit (i.e., $\hbar v_F q\ll E_F$), the density correlation function in the plasmon pole approximation can be obtained purely analytically by (i) substituting Eqs.~(\ref{eq:pop_lowq}) and (\ref{eq:pop_stat}) into Eq.~(\ref{eq:Mermin}), and (ii) using the so obtained $\Pi_\gamma(q,\omega)$ to evaluate the Taylor expansion coefficients $A_q$ and $B_q$ at $\omega=\omega_q$, i.e., where the real part of the denominator of Eq.~(\ref{eq:pop_rpa}) vanishes. The first step produces
\begin{equation}
\Pi_\gamma(q,\omega)\approx\frac{E_F q^2}{\pi\hbar^2 \omega(\omega+i\gamma/\hbar)}.
\end{equation}
The second step produces Eq.~(\ref{eq:plpole}) with
\begin{align}
\Gamma_q&=\gamma, \nonumber\\
\hbar\omega_q&=\sqrt{2 E_F q e^2/\kappa},\nonumber\\
\Lambda_q&=\frac{\hbar^3\omega^3_q}{8\pi E_F}\left(\frac{{\kappa}}{e^2}\right)^2,
\label{eq:plpole_apprx}
\end{align}
thus, resulting in the expected $\omega_q\propto\sqrt{q}$ plasmon dispersion. \cite{Shung1986-979,Wunsch2006-318,Hwang2007-205418}

It turns out that for the specific case considered here, i.e., the plasmon pole approximation in the long wavelength limit, the same analytical expression for $\Pi^{pp}(q,\omega)$ could have been obtained in the limit of small $\gamma$ by using the substitution $\omega\rightarrow\omega+i\gamma/2\hbar$ instead of the more general Mermin's procedure. It has to be emphasized, however, that such an agreement is not general and hard to foresee. Therefore, the more accurate Mermin's procedure has to be favored over more approximate methods of introducing the finite scattering rate into the density correlation function.\cite{Ropke1999-365}

\section{Graphene plasmon in high-$\kappa$ environment}\label{app:high-k}

Plasmon dispersion in graphene remains $\omega_q\propto\sqrt{q}$ only at $q\lesssim \frac{E_F e^2}{{\kappa}\hbar^2 v^2_F}$, and deviates from this simple relation at higher $q$ due to non-local effects originating from interaction of plasmons with intra-band electron-hole pair excitations. Therefore, in high-${\kappa}$ environment, the dispersion of plasmons in graphene can deviate from $\omega\propto\sqrt{q}$ at not very large $q$. To correct for this, we have to find the position of a plasmon pole more accurately than that given by Eq.~(\ref{eq:plpole_apprx}). To this end, we expand the bare density correlation function of graphene at $\hbar\omega,\hbar v_F q\ll E_F$, which yields
\begin{equation}
\Pi_0(q,\omega)=-\frac{2E_F}{\pi\hbar^2 v^2_F}\left[
1-\frac{1}{\sqrt{1-(v_F q/\omega)^2}}
\right].
\end{equation}
It is straightforward to check that at $q\rightarrow 0$ this expression reduces to Eq.~(\ref{eq:pop_lowq}). On the other hand, the limit $\omega\rightarrow 0$ yields Eq.~(\ref{eq:pop_stat}). The equation for the plasmon pole,
\begin{equation}
1-\frac{2\pi e^2}{\kappa q}\Pi_0(q,\omega)=0,
\end{equation}  
is easily solved with respect to $\omega$, yielding the corrected plasmon pole position as
\begin{equation} 
\omega_q=v_F q \frac {\alpha(q)}{\sqrt{\alpha^2(q)-1}},
\end{equation}
where $\alpha(q)=1+\frac{\kappa \hbar^2 v^2_F q}{4e^2 E_F}$. The amplitude of the plasmon pole becomes
\begin{equation}
\Lambda_q=\frac{\kappa \hbar\omega^3_q}{2\pi e^2 v^2_F q}
\frac{1-\sqrt{1-(v_F q/\omega_q)^2}}{1-(v_F q/\omega_q)^2}
\end{equation}
It turns out however that the correction to amplitude is rather small, so the original uncorrected amplitude, given in Eq.~(\ref{eq:plpole_apprx}) can be used.


%

\end{document}